\documentclass[preprint,resetfootnote]{aastex701}
\accepted{May 15, 2026}

\begin{document}

\title{A $\lambda=1.3$ millimeter Survey for Disks around Herbig Be Stars}

\correspondingauthor{David J. Wilner}

\author[0000-0003-1526-7587]{David J. Wilner}
\affiliation{Center for Astrophysics \textbar\, Harvard \& Smithsonian, 60 Garden St., Cambridge, MA 02138, USA}
\email{dwilner@cfa.harvard.edu}

\author[0000-0002-4248-5443]{Joshua Bennett Lovell}
\affiliation{Center for Astrophysics \textbar\, Harvard \& Smithsonian, 60 Garden St., Cambridge, MA 02138, USA}
\email{joshua.lovell@cfa.harvard.edu}

\author[0000-0003-2253-2270]{Sean M. Andrews}
\affiliation{Center for Astrophysics \textbar\, Harvard \& Smithsonian, 60 Garden St., Cambridge, MA 02138, USA}
\email{sandrews@cfa.harvard.edu}

\author[0000-0002-4147-3846]{Miguel Vioque}
\affiliation{European Southern Observatory, Karl-Schwarzschild-Str. 2, 85748 Garching bei München, Germany}
\email{miguel.vioque@eso.org}

\author[0000-0002-7607-719X]{Feng Long}
\affiliation{Kavli Institute for Astronomy and Astrophysics, Peking University, Beijing 100871, People's Republic of China}
\email{long.feng@pku.edu.cn}

\author[0000-0003-4705-3188]{Luca Matr\`a}
\affiliation{School of Physics, Trinity College Dublin, the University of Dublin, College Green, Dublin 2, Ireland}
\email{lmatra@tcd.ie}

\begin{abstract}
We present a survey of 24 Herbig Be stars 
(young stellar objects $>3$~M$_{\odot}$)
within 3~kpc at 1.3~millimeters using the 
Submillimeter Array at $\sim1''$ resolution to identify circumstellar disks and assess 
planet forming potential. 
We detect 1.3~mm emission toward 5 Herbig Be stars that range in mass from 4.3 to 12.9~M$_{\odot}$.
Follow-up observations at 0.87~mm show spectral indices consistent with partly optically thick dust emission.
These millimeter detections are compatible with an extrapolation of the scaling relation 
derived for lower-mass T~Tauri and Herbig~Ae stars between millimeter luminosity and 
stellar host mass, and also millimeter continuum size, suggesting these detections represent 
emission from circumstellar disks. The implied disk masses are sufficient for giant planet formation. 
No decrease in the millimeter detection fraction with stellar host mass is evident within this sample
that would implicate rapid disk dissipation by the radiation fields of the higher mass stars. 
The high fraction of millimeter non-detections is likely due to the survey sensitivity limits 
together with photoevaporation and the dynamical impact of stellar companions.
\end{abstract}


\section{Introduction}
\label{sec:Introduction}

Precision stellar radial velocity studies suggest that the occurrence of giant planets 
peaks around stars with masses near 2~M$_{\odot}$ and drops precipitously above 
about 3~M$_{\odot}$ \citep{Reffert+2015}. A possible cause is that the circumstellar disks 
around higher mass stars are dispersed too quickly to form planets. 
While numerous studies have now shown that the intermediate mass Herbig~Ae stars 
($1.5 <$ M$_* \lesssim3$~M$_{\odot}$) have circumstellar disks similar to the 
lower mass T-Tauri stars, serving as reservoirs for planet formation and conduits for 
mass accretion \citep[e.g.][]{Stapper+2022,Stapper+2025}, there is much less information on the 
properties of disks around young stars of higher masses, the Herbig~Be stars
($3 <$ M$_* \lesssim15$~M$_{\odot}$), \citep[see the review by][]{Brittain+2023}. 

The Herbig~Be stars have received relatively little attention in large part 
because they are intrinsically rare due to the Initial Mass Function,
and they are typically located far away with few examples closer than 1 kpc. 
A particular focus for Herbig~Be star studies has been H$\alpha$ spectropolarimetry,
which probes circumstellar scales of order stellar radii \citep[e.g.][]{Vink+2002}. 
These H$\alpha$ observations identify asymmetries attributed to inner accretion disks,
and show a break in detection rate and depolarization at spectral types of B7$-$B8
($\sim4$~M$_{\odot}$),
suggesting a possible change in the physical mechanism of accretion \citep{Ababakr+2017}.  
Probing  disk material at the larger size scales relevant for planet formation requires 
observations at longer wavelengths sensitive to colder circumstellar material.  
However, these higher mass stars evolve quickly and can remain embedded in the 
cloud material from which they formed, making the relevant circumstellar structures 
difficult to discern unless the observations have very high angular resolution. 
For example, \citet{Sandell+2011} found many Herbig~Be stars they 
observed with the single dish JCMT telescope (14'' beam) were dominated by emission 
from surrounding envelopes, which is not unexpected since the high luminosities of 
these stars can heat large volumes of dust and gas around them.  
The focus of most millimeter/submillimeter high resolution disk studies around 
young B-type stars has been on deeply embedded systems that show sufficiently high 
bolometric luminosity, not on the optically revealed Herbig~Be stars \citep[e.g.,][]{Beltran+2016}. 
At mid-infrared wavelengths, \citet{Verhoeff+2012} probed warm circumstellar dust for a 
sample of Herbig~Be stars with N~band imaging and spectroscopy at subarcsecond resolution,
finding less stellar reprocessing than for Herbig~Ae stars and upper limits on sizes of 
$\sim500$~au, suggesting vertically flat disks, perhaps truncated by photoevaporation.

For the optically revealed Herbig~Be stars, 
in particular those more massive than 4~M$_{\odot}$,
millimeter observations that reach relevant 
size scales for circumstellar disks are both limited and biased to a small number of systems 
located most nearby.  Notably, \citet{Alonso-Albi+2009} observed 6 of the 
closest (0.25 to 1.3 kpc) Herbig~Be stars with the IRAM 
Plateau de Bure Interferometer and/or Very Large Array 
and obtained 4 detections of dust emission, deriving masses at $<1000$~au scales from 
spectral energy distribution modeling typically lower than for Herbig~Ae stars. 
A similar conclusion was obtained from subarcsecond Submillimeter Array (SMA) observations 
of the nearby (0.35 kpc) Herbig~Be star HD~200775 \citep{Okamoto+2009}. 
The prevailing interpretation of these results is the rapid dispersal of disk material 
by the stronger ultraviolet radiation fields of the more massive Herbig~Be stars. 
This scenario finds support in detailed calculations and simulations
of disk photoevaporation \citep{Gorti+2009,Kunitomo+2021,Komaki+2025}. 
Another possibility is tidal stripping of disk material due to high order stellar 
multiplicity at an early evolutionary stage \citep[e.g.,][]{Li+2024}.
Millimeter emission may also be reduced when the radial drift of dust in disks becomes 
very efficient as a high mass star's luminosity dramatically increases 
after a fraction of a Myr \citep{Pinilla+2022}.
Confirmation of any differences between the properties of disks around 
Herbig~Be stars and those around lower mass stars awaits millimeter interferometry 
of a larger sample. A better grasp of disk photoevaporation in the
extreme environments of Herbig~Be stars may also help in understanding the role of 
this potentially important process in the dissipation of disks around lower mass stars. 

The advent of {\em Gaia} has provided the basis for a new and better characterized 
sample of Herbig Be stars. \citet{Vioque+2022} identified Herbig Ae/Be stars using 
optical spectroscopy on candidates selected from a large catalog constructed by 
combining {\em Gaia} data with various photometric and H$\alpha$ surveys 
\citep[using machine learning techniques, see][]{Vioque+2020}. 
This homogeneously selected sample is ideal for characterizing the circumstellar 
environments of the rare Herbig Be stars. 

In this paper, we present new SMA observations of 
24 Herbig Be stars within 3~kpc drawn from the \citet{Vioque+2022} catalog. 
These observations enable a first systematic investigation of cold material 
around forming high-mass stars at resolutions relevant to the scale of circumstellar  disks. 
We describe the target star sample and the SMA observations in \S\ref{sec:Observations}. 
We discuss the resulting photometry, spectral index information,
and the millimeter emission mechanism for the detected Herbig~Be stars in \S\ref{sec:Results}.
We place the millimeter emission from the detections in the context of lower mass stars, 
and address their planet forming potential in \S\ref{sec:Discussion}. 
We present conclusions in \S\ref{sec:Conclusions}.

\section{Observations}
\label{sec:Observations}

\subsection{Herbig~Be star sample}
\begin{figure}[h!]
\centering
\vspace{-1.0truecm}
\includegraphics[width=1.0\textwidth]{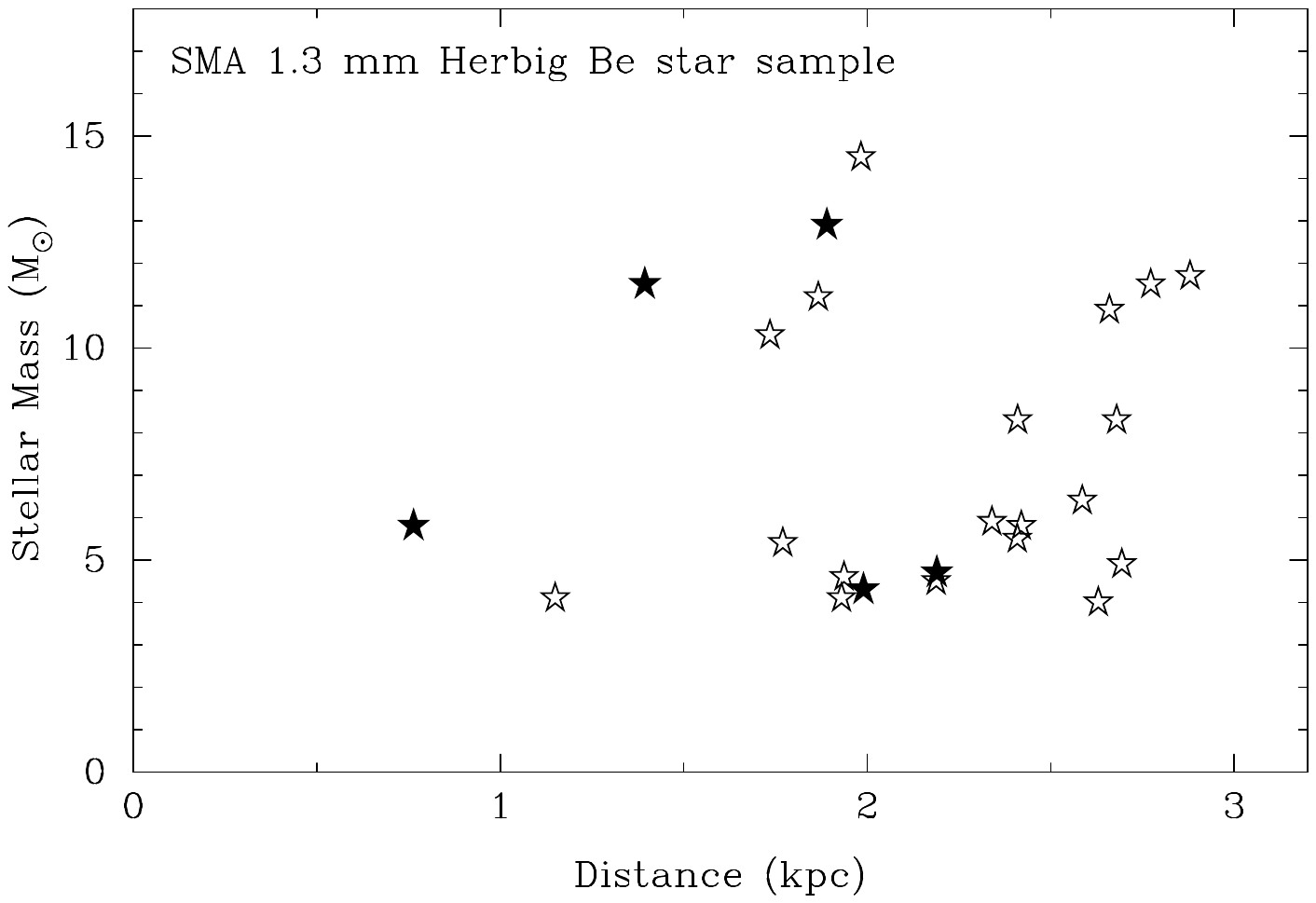}
\vspace{-1.5truecm}
\caption{
The sample of 24 SMA Herbig Be stars targeted by the SMA in the plane of distance and stellar mass.
The distribution samples the full range of Herbig Be stellar masses.
Note that uncertainties in the stellar masses are typically $\pm15$\% but can be larger
(see Table~\ref{tab:sample}). 
The filled symbols denote SMA 1.3~mm detections and the unfilled symbols denote non-detections 
(see \S\ref{sec:Results}).
}
\label{fig:SMA_HBe_sample}
\end{figure}

There are 24 securely identified Herbig Be stars in the \citet{Vioque+2022} catalog
with M$_*$ $>4$~M$_{\odot}$ within 3.0~kpc, and also located north of Dec $-40^{\circ}$ 
that are readily accessible to the SMA. 
Table~\ref{tab:sample} summarizes this Herbig~Be star sample, including 
distances, and estimated stellar masses and ages. 
We caution that these stellar properties, and especially the ages, 
are subject to additional uncertainties from the adopted 
pre-main-sequence tracks \citep[PARSEC V1.2S,][]{Bressan+2012,Marigo+2017}. 
Figure~\ref{fig:SMA_HBe_sample} summarizes the sample in the plane of stellar mass 
and distance, showing that these targets sample the wide range of Herbig~Be star masses 
from $4.0$ to $14.5$~M$_{\odot}$, for systems at distances from $0.8$ to $2.9$~kpc
(with a median distance of 2.2~kpc). 
All of these stars have nominal ages less than 1~Myr and show spectroscopic 
signatures of accretion, as well as infrared excess emission that indicates 
the presence of warm circumstellar dust.
Notably, this sample is also effectively cleaned of the classical Be stars, FS CMa stars, and other 
evolved star contaminants that plague earlier catalogs of purportedly young, massive stars \citep{Vioque+2020}.
While the sample size is modest, these 24 Herbig~Be stars represent nearly 40\% of those known with 
M$_*$ $>4$~M$_{\odot}$ within 3.0~kpc in the expanded {\em Gaia}-based catalog of \citep{Guzman-Diaz+2021}.

\begin{deluxetable}{llclllllll}
%\rotate

\tablecaption{
The target sample of 24 securely identified Herbig~Be stars from the \citet{Vioque+2022} catalog
with M$_* > 4$ M$_\odot$, distance $<3$ kpc, Declination $ >-40\degr$ (accessible to the SMA). 
}
\label{tab:sample}

\tablehead{\colhead{Name} & \colhead{Gaia DR3} & \colhead{Alt Name} & \colhead{RA} & \colhead{DEC} & \colhead{distance} & \colhead{T$_{eff}$} & \colhead{$\log{L_*}$} &  \colhead{M$_*$} & \colhead{Age}  \\  
\colhead{} &     \colhead{} &         \colhead{}          & \colhead{(J2000)} & \colhead{(J2000)} & \colhead{(pc)} & \colhead{(K)} & \colhead{(L$_{\odot}$)} &   \colhead{(M$_{\odot}$)} & \colhead{(Myr)}  } 
\startdata
VOS 70   &  430854252809011200  &  [KW97] 3-17  &  00 36 51.77  &  +63 29 30.1  &  $1150^{+30}_{-28}$  &  $7800^{+200}_{-300}$    & $2.22^{+0.07}_{-0.07}$  &  $~4.1^{+0.35}_{-0.27}$   &  $0.81^{+0.2}_{-0.2}$  \\
VOS 949  &  523968761523650304  &  \nodata  &  00 49 47.54  &  +63 38 10.0  &  $2694^{+76}_{-72}$      &  $9000^{+500}_{-500}$    & $2.58^{+0.11}_{-0.11}$  &  $~4.9^{+0.67}_{-0.55}$   &  $0.5^{+0.24}_{-0.18}$  \\
VOS 738  &  464675226177231744  &  \nodata  &  02 53 59.02  &  +60 39 58.6  &  $1937^{+48}_{-46}$      & $14000^{+500}_{-1500}$   & $2.81^{+0.07}_{-0.16}$  &  $~4.6^{+0.46}_{-0.49}$   &  $0.75^{+0.3}_{-0.21}$  \\
VOS 588  &  466147613983225472  &  GGA 216  &  02 59 05.13  &  +60 54 04.1  &  $2190^{+100}_{-90}$     & $12500^{+1500}_{-1800}$  & $2.77^{+0.18}_{-0.22}$  &  $~4.7^{+1.1}_{-0.8}$     &  $0.67^{+0.57}_{-0.34}$  \\
VOS 934  &  442287245287225856  &  \nodata  &  03 28 32.63  &  +51 13 54.4  &  $1930^{+54}_{-51}$      &  $9000^{+500}_{-500}$    & $2.37^{+0.11}_{-0.10}$  &  $~4.1^{+0.57}_{-0.43}$   &  $0.86^{+0.36}_{-0.3}$  \\
VOS 200  &  250764453016220800  &  \nodata  &  04 05 49.37  &  +51 28 34.5  &  $2420^{+110}_{-100}$    & $10000^{+1000}_{-1000}$  & $2.88^{+0.18}_{-0.20}$  &  $~5.8^{+1.4}_{-1.1}$     &  $0.31^{+0.31}_{-0.17}$  \\
VOS 898  &  258477905042591488  &  \nodata  &  04 30 16.24  &  +48 52 09.9  &  $2188^{+77}_{-72}$      &  $9500^{+750}_{-750}$    & $2.51^{+0.14}_{-0.16}$  &  $~4.5^{+0.81}_{-0.72}$   &  $0.7^{+0.53}_{-0.3}$  \\
VOS 2196  &  205118464010485632  &  MWC 482  &  05 01 20.31  &  +43 32 50.7  &  $2680^{+120}_{-110}$   & $22500^{+2000}_{-2000}$  & $3.54^{+0.15}_{-0.16}$  &  $~8.3^{+1.0}_{-1.1}$       &  $0.21^{+0.69}_{-0.07}$  \\
VOS 595  &  181458215025292160  &  MWC 485  &  05 14 26.92  &  +32 48 03.2  &  $1990^{+130}_{-120}$    & $11000^{+1000}_{-1000}$  & $2.57^{+0.16}_{-0.19}$  &  $~4.3^{+0.87}_{-0.7}$    &  $0.85^{+0.62}_{-0.39}$  \\
VOS 1635  &  3123854268434443648  &  \nodata  &  06 30 27.36  &  +01 44 06.3  &  $2340^{+110}_{-100}$  &  $6000^{+500}_{-500}$    & $2.49^{+0.14}_{-.014}$  &  $~5.9^{+0.6}_{-0.7}$     &  $0.18^{+0.13}_{-0.08}$  \\
VOS 1034  &  4098138462472816384  &  SS 369  &  18 22 44.15  &  $-$15 33 09.0  &  $1735^{+68}_{-63}$   & $20000^{+5000}_{-5000}$  & $3.97^{+0.33}_{-0.42}$  &  $10.3^{+4.5}_{-2.8}$   &  $0.09^{+0.61}_{-0.06}$  \\
VOS 1342  &  4272195138879459200  &  LkHA 348  &  18 34 12.65  &  $-$00 26 21.8  &  $1890^{+100}_{-90}$ & $22500^{+2500}_{-2500}$ & $4.27^{+0.19}_{-0.20}$ &  $12.9^{+2.5}_{-1.9}$   &  $0.052^{+0.032}_{-0.02}$  \\
VOS 1407  &  4155634296310906112  &  \nodata  &  18 37 48.95  &  $-$09 03 34.5  &  $2630^{+210}_{-180}$ & $10400^{+1200}_{-1000}$ & $2.46^{+0.22}_{-0.21}$ &  $~4.0^{+1.1}_{-0.7}$   &  $0.99^{+0.83}_{-0.53}$  \\
VOS 1515  &  4152405172406857088  &  \nodata  &  18 22 00.12  &  $-$13 48 55.6  &  $1983^{+84}_{-77}$  & $28000^{+3000}_{-3000}$  & $4.44^{+0.15}_{-0.20}$  &  $14.5^{+3.2}_{-1.9}$   &  $0.05^{+0.02}_{-0.02}$  \\
VOS 1600  &  4152422554127130240  &  \nodata  &  18 20 58.12  &  $-$13 40 32.4  &  $1867^{+55}_{-52}$  & $22500^{+2500}_{-2500}$  & $4.09^{+0.17}_{-0.18}$  &  $11.2^{+1.8}_{-1.4}$   &  $0.075^{+0.035}_{-0.028}$  \\
VOS 1617  &  4094703381988286592  &  SS 352  &  18 13 13.57  &  $-$19 24 08.5  &  $2660^{+100}_{-90}$  & $24800^{+4200}_{-6300}$  & $4.03^{+0.25}_{-0.40}$  &  $10.9^{+3.4}_{-3.0}$     &  $0.09^{+0.63}_{-0.05}$  \\
VOS 2098  &  4259271891523989376  &  \nodata  &  18 50 31.25  &  $-$01 24 09.6  &  $2410^{+92}_{-86}$  & $14000^{+500}_{-1500}$   & $3.56^{+0.08}_{-0.17}$  &  $~8.3^{+1.0}_{-1.1}$     &  $0.127^{+0.073}_{-0.042}$  \\
VOS 67  &  2031912537697905536  &  V989 Cyg  &  19 48 38.56  &  +30 02 41.4  &  $2586^{+81}_{-76}$     &  $5920^{+120}_{-40}$     & $2.62^{+0.05}_{-0.04}$  &  $~6.4^{+0.18}_{-0.14}$   &  $0.125^{+0.013}_{-0.013}$  \\
VOS 63  &  1836558703328498944  &  \nodata  &  20 10 27.24  &  +27 05 27.7  &  $1770^{+47}_{-44}$      &  $7200^{+240}_{-170}$    & $2.49^{+0.07}_{-0.06}$  &  $~5.4^{+0.32}_{-0.34}$   &  $0.306^{+0.087}_{-0.058}$  \\
VOS 1331  &  2066412811690449792  &  \nodata  &  20 38 45.88  &  +42 07 04.7  &  $1394^{+25}_{-25}$    & $28000^{+3000}_{-3000}$  & $3.98^{+0.13}_{-0.18}$  &  $11.5^{+1.5}_{-1.7}$     &  $0.12^{+0.49}_{-0.05}$  \\
VOS 22  &  2164505844663760768  &  \nodata  &  21 11 19.06  &  +47 38 47.6  &  $2409^{+63}_{-60}$      & $12500^{+2500}_{-2500}$  & $2.98^{+0.25}_{-0.32}$  &  $~5.5^{+2.1}_{-1.4}$     &  $0.41^{+0.62}_{-0.28}$ \\
VOS 2085  &  2007419820986293504  &  LS III+57 89  &  22 47 45.63  &  +58 06 48.8  &  $2880^{+200}_{-180}$ & $24800^{+4200}_{-6300}$ & $4.15^{+0.28}_{-0.42}$  &  $11.7^{+3.6}_{-3.2}$     &  $0.07^{+0.42}_{-0.04}$  \\
VOS 1026  &  2014090042628166656  &  \nodata  &  23 12 26.37  &  +60 58 12.9  &  $2773^{+77}_{-73}$    & $28000^{+3000}_{-3000}$  & $3.97^{+0.13}_{-0.18}$  &  $11.5^{+1.5}_{-1.7}$     &  $0.12^{+0.49}_{-0.05}$  \\
VOS 42  &  2016307791941936896  &  GGR 156  &  23 42 26.73  &  +63 37 38.7  &  $~764^{+10}_{-10}$      & $21000^{+10000}_{-10000}$ & $2.97^{+0.52}_{-0.85}$  &  $~5.8^{+3.2}_{-2.8}$     &  $0.7^{+7.1}_{-0.6}$   \\
\enddata
\end{deluxetable}

\subsection{SMA 1.3~mm observations}
We used the SMA on Maunakea, Hawaii over 4 nights in semester 2023B to observe the 
24 Herbig Be stars in the sample, using the extended (EXT) configuration (baseline lengths
44 to 220~m). The A (230 GHz) and B (240 GHz) receiver sets were each set with 
a Local Oscillator (LO) frequency of 225.5 GHz (1.3~mm). The SWARM digital backend processed an 
IF frequency range of $\pm(4-16)$~GHz for each receiver, for a total continuum 
bandwidth of 48 GHz after combining sidebands and polarizations. 
Table~\ref{tab:SMA_230} provides a log of these 1.3~mm observations, 
including atmospheric opacities, as well as the gain, passband, and flux calibrators.
In each of 4 tracks, observations of 6 targets were interleaved. 
We used the COMPASS 
({\em Calibrator Observations for Measuring the Performance of Array Sensitivity and Stability})
pipeline \citep{Keating2025a} and the 
{\tt pyuvdata} software \citep{Keating2025b} 
to automatically flag spectral channels affected by interference, 
generate calibration tables for bandpass, flux, and time-dependent complex gains,
and to export measurement sets to the CASA software package \citep{CASA+2022}. 
We then used CASA v6.6.3 to apply the calibration tables and the
{\tt tclean} task to make continuum images using robust=2.0 weighting to maximize 
point-source sensitivity.
The estimated uncertainty in the absolute flux scale is about 10\%.
Table~\ref{tab:SMA_230} also lists the resulting beam size and continuum rms noise for each target. 
For these observations, the typical beam size was $\sim1''$ and rms noise $\sim0.63$~mJy.
The SMA primary beam size is $55''$ (FWHM) at 1.3~mm, providing ample field of view for these compact sources. 

\subsection{SMA 0.87~mm observations}
For the 5 Herbig~Be stars in the sample that were detected ($>3\sigma$) at 1.3~mm (see \S\ref{sec:Results}), 
we used the SMA to obtain additional observations in semester 2025A at the shorter wavelength of 0.87~mm.
These follow-up observations were done using the compact (COM) configuration (baseline lengths 20 to 70~m). 
The ``345'' and ``400'' band receivers were each set with an LO frequency of 345.0 GHz, 
providing a total continuum bandwidth of 48~GHz.
Table~\ref{tab:SMA_345} provides a log of these additional observations, 
which were obtained over 2 nights, together with the beam sizes and continuum rms noise.  
The calibration and imaging followed the same procedure as for the 1.3~mm observations.
The typical beam size was $\sim2''$ and rms noise $\sim2$~mJy. 
The SMA primary beam size is $35''$ (FWHM) at this wavelength, again 
much larger than the size scales of the targeted emission structures. 

\begin{deluxetable}{l l c c c c c c} 
\tabletypesize{\footnotesize}
\tablecaption{
Log of the SMA 1.3~mm observations, from the EXT antenna configuration.  
}
\label{tab:SMA_230}

\tablehead{\colhead{Date} & \colhead{Source} & \colhead{$\tau_{225~{\rm GHz}}$} & \multicolumn{3}{c}{Calibrators} & \colhead{beam} & \colhead{rms}\\
\colhead{}    & \colhead{}       & \colhead{}                       & \colhead{gain} & \colhead{passband} & \colhead{flux}  & \colhead{(fwhm, pa)} & \colhead{(mJy)}} 
\startdata
2023 Sep 29 & VOS 2098 & 0.069        & J1830$+$063,J1743$-$038     & 3C84     & Callisto    & $1\farcs26\times0\farcs95,+74\degr$  & 0.56\\  
~           & VOS 67   & \nodata          & J2023$+$318,J2105$+$371   & \nodata   & \nodata  & $1\farcs22\times0\farcs88, -83\degr$ & 0.67 \\  
~           & VOS 63   & \nodata          & J2023$+$318,J2105$+$371   & \nodata   & \nodata  & $1\farcs19\times0\farcs88, -84\degr$ & 0.64 \\  
~           & VOS 1331 & \nodata          & J2015$+$371,MWC349     & \nodata   & \nodata  & $1\farcs18\times0\farcs93, -78\degr$ & 0.63 \\  
~           & VOS 22   & \nodata          & MWC349,J2038$+$513     & \nodata   & \nodata  & $1\farcs19\times0\farcs94, -80\degr$ & 0.54 \\  
~           & VOS 2085 & \nodata          & J0014$+$612,J2202$+$422   & \nodata   & \nodata  & $1\farcs43\times0\farcs99, -82\degr$ & 0.40 \\  
2023 Oct 08 & VOS 70   & 0.108          & J0104$+$612,J0102$+$584    & 3C454.3  & Callisto   & $1\farcs30\times1\farcs14, +85\degr$ & 0.59 \\  
~           & VOS 949  & \nodata          & J0104$+$612,J0102$+$584   & \nodata   & \nodata  & $1\farcs30\times1\farcs14, +86\degr$ & 0.60 \\  
~           & VOS 738  & \nodata          & J0244$+$624,J0359$+$509   & \nodata   & \nodata  & $1\farcs25\times1\farcs12, -88\degr$ & 0.60 \\  
~           & VOS 588  & \nodata          & J0244$+$624,J0359$+$509   & \nodata   & \nodata  & $1\farcs22\times1\farcs13, -81\degr$ & 0.63 \\  
~           & VOS 1026 & \nodata          & J0014$+$612,J0102$+$584   & \nodata   & \nodata  & $1\farcs26\times1\farcs12, -87\degr$ & 0.58 \\  
~           & VOS 42   & \nodata          & J0014$+$612,J0102$+$584  & \nodata   & \nodata  & $1\farcs27\times1\farcs15, -89\degr$ & 0.58 \\  
2023 Oct 10 & VOS 934  & 0.137         & J0346$+$540,J0359$+$509    & 3C454.3  & Callisto    & $1\farcs30\times1\farcs02, -85\degr$ & 0.62 \\  
~           & VOS 200  & \nodata          & J0346$+$540,J0359$+$509   & \nodata   & \nodata  & $1\farcs32\times1\farcs03, -86\degr$ & 0.62 \\  
~           & VOS 898  & \nodata          & J0359$+$509,J0415$+$448   & \nodata   & \nodata  & $1\farcs30\times1\farcs01, -84\degr$ & 0.66 \\  
~           & VOS 2196 & \nodata          & J0423$+$418,J0418$+$380   & \nodata   & \nodata  & $1\farcs30\times0\farcs98, -84\degr$ & 0.64 \\  
~           & VOS 595  & \nodata          & J0418$+$380,J0555$+$398   & \nodata   & \nodata  & $1\farcs28\times0\farcs94, -85\degr$ & 0.63 \\  
~           & VOS 1635 & \nodata          & J0607$-$085,J0725$-$009   & \nodata   & \nodata  & $1\farcs25\times0\farcs98, +85\degr$ & 0.66 \\  
2024 Apr 05 & VOS 1617 & 0.058         & J1832$-$105,J1733$-$130    & 3C273  & Ceres         & $1\farcs17\times1\farcs15, +84\degr$ & 0.80 \\  
~           & VOS 1600 & \nodata          & J1832$-$105,J1733$-$130   & \nodata   & \nodata  & $1\farcs25\times0\farcs98, +85\degr$ & 0.73 \\  
~           & VOS 1515 & \nodata          & J1832$-$105,J1733$-$130   & \nodata   & \nodata  & $1\farcs19\times1\farcs06, -75\degr$ & 0.68 \\  
~           & VOS 1034 & \nodata          & J1832$-$105,J1733$-$130   & \nodata   & \nodata  & $1\farcs27\times1\farcs07, -68\degr$ & 0.76 \\  
~           & VOS 1342 & \nodata          & J1832$-$105,J1733$-$130   & \nodata   & \nodata  & $1\farcs24\times0\farcs96, -86\degr$ & 0.68 \\  
~           & VOS 1407 & \nodata          & J1832$-$105,J1733$-$130   & \nodata   & \nodata  & $1\farcs27\times1\farcs01, -78\degr$ & 0.70 \\  
\enddata
\end{deluxetable}

\begin{deluxetable}{l l c c c c c c} 
\tabletypesize{\footnotesize}
\tablecaption{
Log of the SMA 0.87~mm observations, from the COM antenna configuration.  
}
\label{tab:SMA_345}
\tablehead{\colhead{Date} & \colhead{Source} & \colhead{$\tau_{225~{\rm GHz}}$} & \multicolumn{3}{c}{Calibrators} & \colhead{beam} & \colhead{rms} \\
\colhead{}    & \colhead{}       & \colhead{}                       & \colhead{gain} & \colhead{passband} & \colhead{flux}  & \colhead{(fwhm, pa)} & \colhead{(mJy)}} 
\startdata
2025 Aug 01 & VOS 588 & 0.061         & J0244$+$624,J0359$+$509     & 3C84     & Vesta & $2\farcs92\times1\farcs79,+51\degr$  & 2.09\\  
~           & VOS 1342   & \nodata       & J1743$-$038,J1830$+$063  & \nodata  & \nodata  & $2\farcs29\times2\farcs14, -69\degr$ & 2.14 \\  
~           & VOS 1331 & \nodata         & J2015$+$371,MWC349       & \nodata  & \nodata  & $2\farcs46\times1\farcs78, +54\degr$ & 1.75 \\  
~           & VOS 42   & \nodata         & J0014$+$612,J0102$+$584  & \nodata  & \nodata  & $2\farcs81\times1\farcs82, +41\degr$ & 2.48 \\  
2025 Aug 05 & VOS 595   & 0.108          & J0418$+$380,J0555$+$398  & 3C84     & Uranus & $2\farcs11\times1\farcs91, -81\degr$ & 2.69 \\  
\enddata
\end{deluxetable}

\section{Results}
\label{sec:Results}

The SMA 1.3~mm observations resulted in 5 detections above the $3\sigma$ level at the targeted stellar
positions, for a detection rate of 5/24 = 21\%.   
These detections include some of the lowest mass and the highest mass Herbig~Be  stars 
in the sample (indicated by the filled symbols in Figure~\ref{fig:SMA_HBe_sample}).
These detections also include some of the youngest and oldest stars in the sample.
The fact that all of the detections are for stars closer than 2.2~kpc perhaps suggest a 
bias due to distance, although the decrease in millimeter luminosity 
sensitivity from 2.2 to 3.0 pc is only about a factor of two. 

\subsection{SMA images and photometry}
\begin{figure}
\centering
\includegraphics[width=1.0\textwidth]{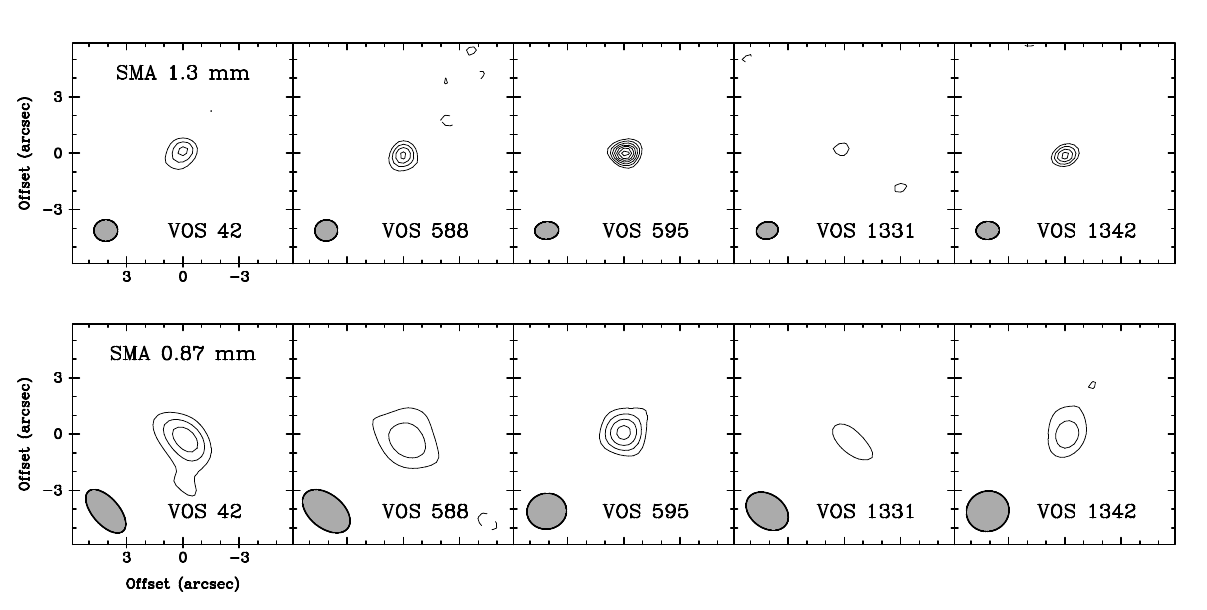}
\caption{SMA images of the Herbig Be stars that were significantly detected at 1.3~mm (top)
and 0.87~mm (bottom). The ellipses in the lower left corner of each panel show the beams,  
and the contour levels are $3,6,9,...\times$ rms noise in each image (see Table~\ref{tab:Results}).
\label{fig:SMA_detections}
}
\end{figure}

Figure~\ref{fig:SMA_detections} shows continuum images of the Herbig Be systems that were detected 
at 1.3~mm and followed up at 0.87~mm. The emission from each source appears compact, and
comparable to the beam size. 
Point source and Gaussian model fits to the visibilities and to the images suggest that only 
VOS 42 -- the closest source in the sample -- was spatially resolved in the higher resolution 1.3~mm observations, 
albeit very marginally, with a deconvolved (fwhm) size of $1\farcs0\pm0\farcs4$ ($760\pm300$~au). 
Table~\ref{tab:Results} summarizes the photometry of these sources; the reported values represent
point source model fits except for VOS 42 for which emission was integrated in a $\pm3\farcs0$ box. 
The uncertainties include  a 10\% systematic error in the flux scale for each band 
added in quadrature to the statistical uncertainties.

Table~\ref{tab:Results} also includes results for the emission extent for each source. 

\begin{deluxetable}{l c c c c c c} 
\tablecaption{
Observed and derived properties for the disks around the Herbig~Be stars detected by the 
SMA at 1.3~mm and 0.87~mm, 
including flux measurements, spectral indices between these wavelengths, disk masses. 
and size constraints.
The disk masses assume optically thin emission and a standard mass opacity and temperature
(see S\ref{sec:Discussion-masses}). 
The upper limits for disk radii correspond to half of the 1.3~mm synthesized beam size 
except for VOS~42 where the radius corresponds to half of the deconvolved size from a 
Gaussian fit.
}
\label{tab:Results}
\tablehead{\colhead{Source} &  \colhead{F$_{1.3~{\rm mm}}$} &  \colhead{F$_{0.87~{\rm mm}}$} & \colhead{$\alpha_{mm}$\tablenotemark{a}} & \colhead{M$_{disk}$}      & \colhead{M$_{disk}$/M$_*$} & \colhead{R$_{disk}$} \\
\colhead{}    & \colhead{(mJy)}       & \colhead{(mJy)}                       & \colhead{} & \colhead{(M$_{\odot}$)} & \colhead{}  
& \colhead{($''$)}}  \

\startdata
VOS 588   & $ 6.3\pm0.9$ & $17.8\pm2.7$ & $2.44\pm0.49$ & 0.091 & 0.019  & $<0.59$ ($1290$~au) \\ 
VOS 1342  & $ 6.3\pm0.9$ & $14.1\pm2.5$ & $1.90\pm0.55$ & 0.068 & 0.0053 & $<0.55$ ($1030$~au) \\
VOS 1331  & $ 2.7\pm0.7$ & $ 8.4\pm2.0$ & $2.67\pm0.80$ & 0.016 & 0.0014 & $<0.53$ ($730$~au)  \\
VOS 42    & $ 7.4\pm0.9$ & $22.7\pm3.4$ & $2.64\pm0.46$ & 0.013 & 0.0022 & $0.50\pm0.20$ ($380\pm150$~au) \\
VOS 595   & $10.2\pm1.2$ & $26.5\pm3.8$ & $2.25\pm0.43$ & 0.121 & 0.028  & $<0.56$ ($1090$~au) \\
\enddata
\tablenotetext{a}{spectral index between 1.3 and 0.87~mm %, i.e. $F_\nu\propto\nu^{\alpha_{mm}}$ 
}
\end{deluxetable}

\subsection{Millimeter emission mechanism}
The millimeter emission from circumstellar environments of T Tauri and 
Herbig Ae stars generally can be safely attributed to dust, with rare 
exceptions due to powerful ionized winds or non-thermal stellar activity.
The higher mass Herbig~Be stars have much stronger ultraviolet radiation 
fields, however, which can ionize gas near the star and 
produce bremsstrahlung (free-free emission) that could contribute -- or even dominate-- the 
emission at millimeter wavelengths. This is especially a concern for the
stars with earlier spectral types, as the ultraviolet radiation fields
increase substantially at the higher effective temperatures of these stars. 
As an example, \citet{Alonso-Albi+2009} analyzed the well sampled radio-to-millimeter 
spectrum of MWC~137, a 14~M$_{\odot}$ Herbig~Be star (or perhaps higher mass, 
see \citealt{Kraus+2021}), and found a  compact component with spectral index 
$\alpha =0.79$ (i.e. $F_\nu\propto\nu^{\alpha}$) 
indicative of emission from partly optically thick ionized gas that 
dominates at 1.3~mm. The MWC~137 system is also clearly visible 
in VLA Sky Survey (VLASS) images \citep{Lacy+2020} at $10$~cm, a sufficiently 
long wavelength that circumstellar dust emission would not be detectable. 

We can address, in part, concerns about significant contributions 
from ionized gas at 1.3~mm for the Herbig Be stars in the SMA sample
using (1) VLASS 10~cm images, 
and (2) the spectral index observed in the 1.3~mm to 0.87~mm 
wavelength range.

We downloaded the fits format VLASS quicklook images for the 
available epochs and the median quicklook images\footnote{https://science.nrao.edu/vlass}.
We found no detections at the source positions within the $\sim3''$ VLASS beam, 
and rms noise levels commensurate with the expected values of $\sim80~\mu$Jy~beam$^{-1}$.
Given the near-flat spectral index of optically thin free-free emission 
($\alpha = -0.1$), or the decreasing spectral index from 
synchrotron emission ($\alpha = -0.7$ for a standard spectrum),
emission from these mechanisms would fall well below the 
rms noise levels of the SMA 1.3~mm observations ($400$ to $800$~$\mu$Jy). 
However, for partly optically thick ionized gas with a rising spectrum, 
e.g. a spherical ionized wind ($\alpha = 0.6$), or the even steeper 
spectrum of a {\em collimated} ionized wind  \citep{Reynolds1986},
the signal could lie below the VLASS detection limit and still 
contribute significantly at 1.3~mm. Specifically, emission from ionized gas 
with a spectral index $\alpha=0.6$ (or $1.0$) that remains undetected 
in the VLASS images could contribute up to $\sim1$ (or 6)~mJy at 1.3~mm. 

Disks around T-Tauri and Herbig~Ae stars 
typically show $\alpha= 2.2\pm0.3$ from 1.3~mm to 0.87~mm \citep{Andrews2020}.
These spectral index values are interpreted as a mix of optically thick dust emission
and optically thin dust emission with a steeper spectral index that depends 
in detail on grain sizes and composition \citep[e.g. $\alpha \approx 3.0$,][]{Beckwith+1990}.
As listed in Table~\ref{tab:Results}, the spectral index values 
for the Herbig Be stars in the SMA survey
range from 1.9 to 2.7; these values are all consistent 
with dust emission, given the uncertainties, and consistent 
with the typical values for dusty disks around lower mass stars.

Taken together, the lack of VLASS detections and millimeter spectra consistent
with dusty disks suggests that dust emission dominates for the detected systems. 
Of course, more complicated scenarios that combine multiple emission components,
e.g. dust emission and partly optically thick ionized gas cannot be ruled out 
with the data available.

\section{Discussion}
\label{sec:Discussion}
We surveyed a sample of 24 Herbig~Be stars using millimeter interferometry to probe 
stellocentric radial scales of 500 to 1500~au, and detected 5 of them. The millimeter spectral indices 
of these detected sources are consistent with dust emission. We next examine if the millimeter
emission is compatible with dusty circumstellar disks based on extrapolations of the empirical 
scaling relations for disk millimeter luminosities observed for lower mass stars. 
We also discuss the implied circumstellar disk masses, their planet forming potential, 
and speculate about the high fraction of non-detections. 

\subsection{Millimeter Scaling Relations for Protoplanetary Disks}
\label{sec:Discussion-trends}
\subsubsection{Millimeter luminosity}
Surveys of millimeter continuum luminosities from T Tauri and Herbig Ae stars
show that circumstellar disk mass increases with stellar host mass, up to about $3$~M$_{\odot}$, 
albeit with significant scatter \citep{Andrews+2013,Andrews2020}. 
At 1.3~mm, the (mean) scaling relation between millimeter luminosity and 
stellar host mass can be expressed approximately as 
$\bar{L}_{{\rm 1.3~mm}} ({\rm mJy}) \approx 40~({M_*}/{1~{\rm M_{\odot}}})^{1.7} ({\rm d}/{150~{\rm pc}})^{-2}.$
If an extrapolation to higher stellar masses holds, then a 4.5~M$_{\odot}$ 
(or 10~M$_{\odot}$) star at 2~kpc would host a disk of about 3 (or 11)~mJy.
As Table~\ref{tab:Results} shows, the handful of SMA 1.3~mm {\em detections} of Herbig~Be systems
are broadly consistent with these extrapolations of the empirical correlation. 

\begin{figure}[h!]
\centering
\vspace{-1.0truecm}
\includegraphics[width=1.0\textwidth]{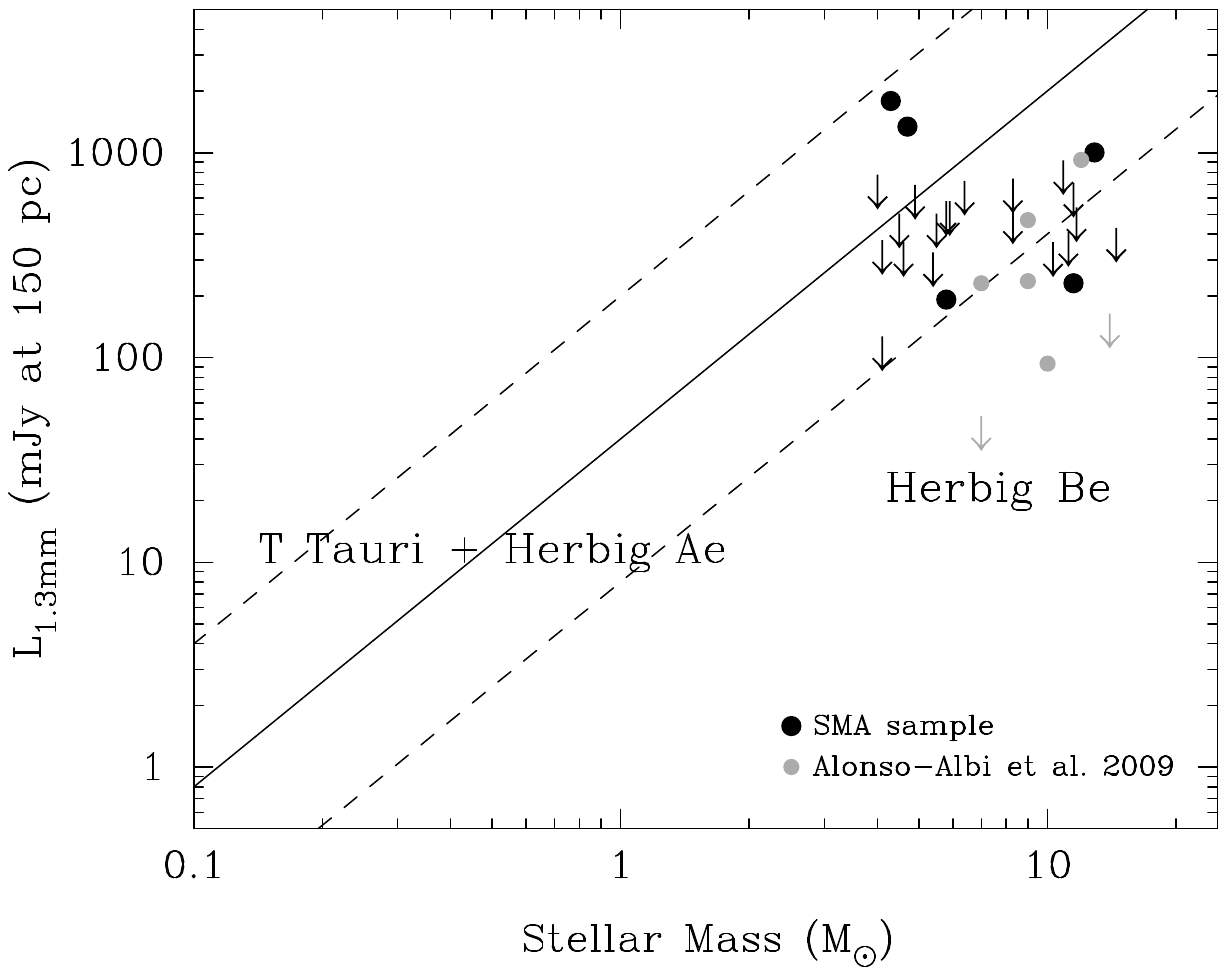}
\vspace{-1.5truecm}
\caption{The solid line traces the mean scaling relation between millimeter luminosity 
and stellar host mass, and the dotted lines indicate $\pm0.7$ dex intrinsic scatter
\citep{Andrews+2013}. 
SMA 1.3~mm detections and upper limits for Herbig Be stars are indicated by the black
circles and downward arrows, respectively. Additional results for Herbig~Be stars 
in grey are from \citet{Alonso-Albi+2009}. 
}
\label{fig:Lmm_Mstar}
\end{figure}

Figure~\ref{fig:Lmm_Mstar} shows the stellar host mass scaling relation extrapolated 
to higher masses, with the 24 Herbig~Be stars observed by the SMA together with the 
6 Herbig Be stars observed by \citet{Alonso-Albi+2009} (after correction for emission
from ionized gas).  
As noted in \S\ref{sec:Introduction}, previous millimeter observations of Herbig~Be stars 
hint at a deficit in disk mass for Herbig~Be stars compared to Herbig~Ae stars \citep{Alonso-Albi+2009}.
As expected, these observations all fall below the extrapolation of the mean scaling relation.
By contrast, the new millimeter detections are generally in line with the extrapolation of the correlation
found for lower mass stars, considering the intrinsic $\pm0.7$~dex scatter in that relation.
Given the combination of the noise level and the intrinsic scatter, 
some non-detections would be expected for the SMA observations at the distances of the sample. 
However, the high non-detection rate for millimeter emission of $\sim80$\% suggests that a simple 
extrapolation of the scaling relation to Herbig~Be stars probably does not apply.

One possible explanation for the non-detections is the effects of stellar companions 
on the Herbig Be star disks in the sample. The scaling relation for lower mass stars excludes 
known binaries, since the millimeter luminosities for such stars is found to be depressed, 
especially when the binary separation is on the cirumstellar disk scale,
presumably due to dynamical interactions and tidal truncation of disks \citep{Harris+2012,Akeson+2019}.
We do not have detailed constraints on the multiplicity of the Herbig~Be stars in 
the \citet{Vioque+2022} sample, but the {\em Gaia} catalog 
goodness-of-fit indicators for single-star models suggest that many of them may have
stellar companions, at least at separations comparable to the {\em Gaia} resolution 
of 0\farcs5. Specifically, VOS~1034, VOS~1342, VOS~1515, VOS~200, VOS~2098, VOS~42, VOS~63, 
and VOS~70 are flagged by the {\tt idp\_gof\_harmonic\_amplitude} indicator, 
implying elongated images that may be associated with partially resolved companions,
and VOS~588 and VOS~898 by {\tt idp\_frac\_multi\_peak} consistent with
two barely resolved sources  (see \citealt{Lindegren+2021} for descriptions of these indicators). 
As the SMA detections include 3 sources with these indicators
VOS~1342, VOS~42, VOS~588), 
they do not preclude 
the presence of circumstellar disks. 
If these indicators are due to the presence of stellar companions, then the orbital separations 
may simply be too wide to have a significant dynamical influence. 

Population statistics also suggest that a significant fraction of the sample should 
have stellar companions. Observations of Herbig~Be stars using a variety of techniques 
show a multiplicity fraction as high as 70\% at wide separations of 100's of au 
\citep{Baines+2006,Wheelwright+2010}. 
Direct imaging shows that companions within 100~au are rare for 
Herbig~Ae/Be stars, and companions around Herbig~Be stars tend 
to be more distant than those around Herbig~Ae stars \citep{Thomas+2023}.
This difference is likely affected by observational limitations, as 
higher contrast is required to detect close companions around Herbig~Be stars.

On the other hand, the statistics for {\em main-sequence} B-type stars show a 
high multiplicity frequency that increases with primary mass 
from $81\pm6$\% for $3 - 5$~M$_{\odot}$ to $93\pm4$\% for $8 - 17$~M$_{\odot}$,
with an orbital distribution uniform in $\log{(a)}$ for semi-major axis $a = 0.1 - 1000$~au \citep{Moe+2017,Offner+2023}. 
Notably, radial velocity studies of main-sequence B-type stars show a 
multiplicity fraction that increases from about 15\% for B9 stars to 60\% for B0 stars 
\citep{Chini+2012}. Since the orbital separations of radial velocity companions overlap 
with circumstellar disk scales, tidal disruption due to such companions, if present, is a 
plausible explanation for at least some of the SMA non-detections. Multi-wavelength observations 
with  higher resolution and sensitivity will be needed to directly address this issue. 

Observations by \citet{Vioque+2018} show a systematic decline in near-infrared excess among Herbig~Be stars 
with masses $>7$~M$_{\odot}$, suggesting that the intense ultraviolet radiation from these more massive stars 
may efficiently clear their hot inner disks. In contrast, the SMA sample does not exhibit a corresponding 
trend with stellar mass: detections were made for 3 out of 14 stars $<7$~M$_{\odot}$ and 2 out of 10 for stars $>7$~M$_{\odot}$. 
Since these subsamples span similar distances, i.e., the higher-mass stars are not preferentially 
more distant (see Figure \ref{fig:SMA_HBe_sample}), this lack of trend is unlikely to be a sensitivity effect. 
The circumstellar material in the outer disk (traced by millimeter emission) shows no evidence for a stellar-mass 
dependent decline like the inner disk (traced by near-infrared excess emission). 

It is possible that the timescale for clearing the outer disk plays a role 
in the difference between the inner and outer disk trends with stellar mass.
Full disk dispersal by photoevaporation around 7~M$_{\odot}$ (and higher) stars is 
predicted to be rapid, only $\sim10^5$ years in the calculations by \citet{Gorti+2009} and 
$7\times10^{6}$ years in the simulations of \citet{Komaki+2025}. 
These dispersal timescales in the models are comparable to, or longer than, the 
age estimates of $0.5\times10^{5}$ and $1.2\times10^{5}$ years for the two stars 
$>7$~M$_{\odot}$ detected in the sample (VOS~1342 and VOS~1331). It is plausible 
that photoevaporation is underway around these two massive stars, but incomplete. 
This idea is consistent with the fact that the younger of the two massive stars 
has a higher disk luminosity and therefore higher disk mass (see S\ref{sec:Discussion-masses}),
although this could also be due to the intrinsic spread in initial disk masses. 
Other stars in the sample with similar high masses and young ages were 
not detected.  A caveat in this comparison to photoevaporation models is that the stellar 
isochronal ages suffer from systematic uncertainties in the pre-main-sequence evolutionary 
tracks  \citep{Vioque+2022}.

\subsubsection{Millimeter emission size}
Demographic surveys of resolved disks around lower mass stars show a 
scaling relation where the millimeter luminosity appears to be proportional to 
the effective emitting area, i.e. $L_{mm}\propto R_{mm}^2$, where $R_{mm}$ is 
an empirical disk size measure \citep{Tripathi+2017,Andrews+2018}.
The millimeter size constraints from the SMA observations of the detected
Herbig~Be stars can be compared to these expectations. 
Taking the best-fit scaling relation at face-value from \citet{Andrews+2018}
implies that a disk with a characteristic flux $F_{\rm 1.3mm} = 6$~mJy 
at 2~kpc distance would have an effective radius of about 140~au, 
i.e. approaching the largest disk radii observed around T~Tauri and Herbig~Ae stars,
a few of which are known to exceed 200~au \citep{Stapper+2022}.
This effective disk radius corresponds to an angular size of $0\farcs07$ on the sky
at a distance 2~kpc, consistent with the emission appearing unresolved in the typical 
$1''$ beam size used for these SMA survey observations.

In this context, it is notable that VOS~42 -- the one source that appears
marginally resolved by the SMA -- is the closest one in the sample (0.764~kpc). 
The scaling relation implies an effective radius of about 60~au, 
and the presence of a disk of this radius, consistent with the 
marginally resolved size 
of $380\pm150$~au ($0\farcs5\pm0\farcs2$, see Table~\ref{tab:Results}), 
might explain why the emission does not appear purely point-like, 
although the uncertainties are very large. 
Alternatively, extended millimeter emission could arise from the inner region 
of an envelope component in this system. 
VOS~42 also shows a relatively steep spectral index from among the
SMA detections, which could indicate a contribution to the emission 
from the inner regions of an extended and optically thin envelope. 

\subsection{Planet formation potential}
\label{sec:Discussion-masses}
High-mass stars present an extreme environment for planet formation, and planet detection around 
B-type stars presents challenges for standard techniques such as radial velocities and transits. 
Consequently, determining properties of the disks 
around the Herbig Be stars can provide a useful, if indirect, way to constrain their 
potential to form planets. A fundamental issue is whether or not the disks around Herbig~Be stars
have sufficient mass for giant planet formation, either by core accretion or gravitational instability.

Thermal continuum emission from dust at millimeter wavelengths provides the best observational 
proxy for total disk mass. 
Following \citet{Beckwith+1990}, we can translate the measurements of disk scale millimeter 
continuum emission into estimates of disk mass (gas + dust), making the standard assumptions 
that the dust is optically thin and isothermal, i.e. 
\begin{equation}
M_{disk}  =  \frac{F_\nu d^2}{\kappa_\nu B_\nu(T)}  
           \approx 0.024~M_{\odot} \left(\frac{F_{\rm 1.3mm}}{2~{\rm mJy}}\right) 
                                     \left(\frac{d}{2.0~{\rm kpc}}\right)^2 
                                     \left(\frac{50~{\rm K}}{\langle T \rangle}\right)
\end{equation}
\noindent
where $F_{\rm 1.3~mm}$ is the 1.3~mm~flux density, $d$ is the distance, 
$\kappa = 0.02$~cm$^2$~g$^{-1}$ is the mass opacity at 1.3~mm 
(including an assumed gas-to-dust ratio of 100), 
and $B_\nu(\langle T \rangle)$ is the Planck function taken to be in the 
Rayleigh-Jeans regime for an appropriate average disk temperature ${\langle T \rangle}$.
While these standard assumptions are known to be imperfect and optical depth
is no doubt important \citep[e.g.,][]{Zhu+2019}, they facilitate comparison with all other 
circumstellar disk demographic surveys based on millimeter continuum emission \citep[e.g.][]{Manara+2023}.
For context, the $3\sigma$ detection limit for the SMA observations of F$_{\rm 1.3mm} \approx 2$~mJy 
corresponds to $\sim25$~Jupiter masses of gas+dust at the $2.0$~kpc,
an amount relevant for forming the systems of super-Jovian mass planets that 
have been detected around A-type stars such as $\beta$~Pictoris and HR~8799.
Table~\ref{tab:Results} lists the disk masses from the SMA 1.3~mm detections. 
These lie in the range 13 to 130 Jupiter masses,
indicating that at least some Herbig~Be systems contain sufficient disk mass 
to form giant planets.

For these standard assumptions about the conversion from dust emission to mass, 
\citep{Andrews+2013} find that disk mass scales approximately linearly 
with stellar host mass for T~Tauri and Herbig~Ae stars, 
with a typical disk-to-star mass ratio of about 0.5\%. 
This correlation between disk mass and stellar mass has been suggested to underlie the 
correlation between giant planet frequency and stellar host mass, in the observed range.
Table~\ref{tab:Results} lists the nominal ratios of disk mass to stellar mass for 
these Herbig~Be systems. These ratios lie in the range $\approx0.1$ to 3\%, in line with 
the ratios found for disks around lower mass stars. 
More sensitive millimeter observations are needed to show if there are additional systems 
in this sample that have giant planet forming potential. 

The (small) fraction of Herbig~Be stars with such massive circumstellar disks can be 
compared with inferences for the fraction of B stars that harbor planetary mass companions. 
High contrast imaging of 42 B stars in the nearby $\sim16$~Myr-old Sco-Cen 
association in the B-star Exoplanet Abundance Study (BEAST) survey found a substellar 
detection rate of $11^{+7}_{-5}$\% at separations of 10 to 1000 au \citep{Delorme+2024}.
As these directly imaged companion detections from BEAST have large semi-major axes 
of 560 and 290 au, and their masses straddle the deuterium burning limit, it is not
entirely clear whether or not they formed from circumstellar disks.
Another comparison can be made to the fraction of white dwarfs with metal pollution 
-- a sign of planets. 
This fraction drops precipitously from $44\pm6$\% for progenitor stars with masses 
less than about 3.6~M$_{\odot}$ 
to $11^{+6}_{-4}$\% for progenitor stars of higher mass \citep{OuldRouis+2024}.  
A single candidate giant planet companion in a sample of young, massive
white dwarfs identifed through infrared excess suggests a similar
giant planet occurrence fraction of $11^{+13}_{-7}$\% 
for stars with initial masses $\gtrsim3$~M$_{\odot}$ \citep{Cheng+2025}.
A precision radial velocity survey that included 113 evolved giant stars in the mass range 
2.7 to 5~M$_{\odot}$ found no planets at all in this host mass regime \citep{Reffert+2015}. 
Taken together, the evidence for giant planets around main-sequence B stars seems
compatible with the detection fraction of disks around Herbig~Be stars, although both would
clearly benefit from better statistics.  Given that the radial velocity observations 
are sensitive primarily to companions with close separations, the lack of planet
candidates from this technique may suggest rapid dispersal of the inner disks may 
interfere with the planet formation process, or perhaps with subsequent planet migration.

\subsection{Limitations and Prospects}
While this SMA survey detected dust emission on $<1000$~au scales in $\sim20$\% of 
the sample of Herbig~Be stars, deeper millimeter continuum observations are needed to 
access the full range of disk masses likely to be present in the Herbig~Be population.
In addition, sensitive observations at longer wavelengths are needed to constrain 
the possible contributions of free-free emission in the millimeter. 

Moreover, while the detected millimeter emission is consistent with dusty disks, observations with 
an order of magnitude higher angular resolution are needed to confirm disk morphologies. 
Such higher resolution imaging also has the potential to directly reveal 
differences in the disk structure, such as inner disk clearing due to incomplete disk photoevaporation. 
Future work should include resolved imaging of CO line emission to confirm 
Keplerian disk kinematics, and to determine full gas extents and viewing geometries.  
High resolution observations of CO emission from R~Mon, 
one of the closest Herbig~Be stars at 0.8 kpc, directly resolves Keplerian rotation and
provides a proof-of-concept for such studies \citep{Fuente+2006,Alonso-Albi+2018}. 
In addition, observations of rare CO isotopologues can provide alternative constraints 
on the masses of circumstellar structures. Finally, observations of a 
larger sample of Herbig~Be stars are desirable to better constrain the population statistics, 
and to address the possibility of trends related to stellar mass and age. 
Because the SMA survey included nearly 40\% of the Herbig~Be stars identified 
within 3~kpc (with M$_*>4$~M$_{\odot}$), a material improvement in the statistics 
will likely require observing systems located at greater distances.

Nonetheless, all of these objectives are well within the observational capabilities 
the Atacama Large Millimeter/submillimeter Array together 
with the Karl G. Janksy Very Large Array.

\section{Conclusions}
\label{sec:Conclusions}
We used the SMA to make 1.3~mm observations of 24 Herbig~Be stars 
(young stellar objects $>3$~M$_{\odot}$) selected from 
the \citet{Vioque+2022} catalog.
The $\sim$1'' angular resolution of these observations probes stellocentric radii $\lesssim1000$
for these systems, and provides a first step to characterize the millimeter emission
at these size scales around Herbig~Be stars in a systematic way.
The main results are:
\begin{enumerate}
\item Millimeter emission was detected toward 5 Herbig~Be star with stellar masses
that range from 4.3 to 12.9 M$_{\odot}$. 
\item The spectral indices of the detected emission between 1.3~mm and 0.87~mm lie 
within the range typically seen in disks around lower-mass stars; this is consistent 
with a mix of optically thick and thin dust emission, but may also reflect a mix of dust 
and partly optically thick free-free emission.
\item The detected millimeter emission is consistent with extrapolations of 
empirical scaling relations between disk millimeter luminosity 
and stellar host mass, and also disk millimeter extent.
The closest source, VOS 42, appears marginally resolved.  
\item Interpreting the millimeter detections as emission from circumstellar disks, 
the implied masses are sufficient to form giant planets (adopting standard assumptions). 
\item The $\sim80$\% fraction of non-detections in this millimeter survey is likely 
explained by a combination of the sensitivity limits of the 
observations, the presence of multiple star systems with truncated or disrupted circumstellar disks, 
and the action of photoevaporation. 
\end{enumerate}
For the detected sources, deeper millimeter observations with higher angular resolution 
are needed to confirm disk morphologies, search for disk substructures associated with ongoing 
photoevaporation, and demonstrate gas in Keplerian rotation.
For the non-detections, deeper observations are needed to determine if they represent
a population of fainter (and smaller) disks, and if there are trends with stellar
host mass and/or age indicative of disk photoevoporation.  

\begin{acknowledgments}
The Submillimeter Array is a joint project between the Smithsonian Astrophysical Observatory 
and the Academia Sinica Institute of Astronomy and Astrophysics and is funded by the 
Smithsonian Institution and the Academia Sinica.
We recognize that Maunakea is a culturally important site for the indigenous Hawaiian people; 
we are privileged to study the cosmos from its summit.
J.B.L. acknowledges the Smithsonian Institute for funding via the Center for Astrophysics JC Ryan Fellowship, 
and the Submillimeter Array (SMA) Fellowship.
\end{acknowledgments}

\vspace{5mm}
\facilities{SMA}

%\software{astropy \citep{2013A&A...558A..33A,2018AJ....156..123A},  
          
%\bibliography{HerbigBe.bib}{}
%\bibliographystyle{aasjournalv7}

\end{document}